\documentclass[%
superscriptaddress,
preprint,
floatfix,
amsmath,
amssymb,
pra,
]{revtex4-2}

\usepackage{graphicx}% Include figure files
\usepackage{dcolumn}% Align table columns on decimal point
\usepackage{bm}% bold math
\usepackage{longtable}
\usepackage{booktabs}
\usepackage{multirow}
\usepackage[normalem]{ulem}
\usepackage{xcolor}
\UseRawInputEncoding

\begin{document}

\title{Ab initio MCDHF calculations of the In and Tl electron affinities \\ and their isotope shifts}
\author{Ran Si}
\affiliation{Shanghai EBIT Lab, Key Laboratory of Nuclear Physics and Ion-beam Application, Institute of Modern Physics, Department of Nuclear Science and Technology, Fudan University, Shanghai 200433, Peoples Republic of China}
\affiliation{Spectroscopy, Quantum Chemistry and Atmospheric Remote Sensing (SQUARES), CP160/09, Universit\'{e} libre de Bruxelles, 1050 Brussels, Belgium}
\author{Sacha Schiffmann}
\affiliation{Spectroscopy, Quantum Chemistry and Atmospheric Remote Sensing (SQUARES), CP160/09, Universit\'{e} libre de Bruxelles, 1050 Brussels, Belgium}
\affiliation{Division of Mathematical Physics, Department of Physics, Lund University, Box 118, SE-22100 Lund, Sweden}
\author{Kai Wang}
\affiliation{Hebei Key Lab of Optic-electronic Information and Materials, The College of Physics Science and Technology, Hebei University, Baoding 071002, Peoples Republic of China}
\author{Chong Yang Chen}%
\email{chychen@fudan.edu.cn}
\affiliation{Shanghai EBIT Lab, Key Laboratory of Nuclear Physics and Ion-beam Application, Institute of Modern Physics, Department of Nuclear Science and Technology, Fudan University, Shanghai 200433, Peoples Republic of China}
\author{Michel Godefroid}
\email{Michel.Godefroid@ulb.be}
\affiliation{Spectroscopy, Quantum Chemistry and Atmospheric Remote Sensing (SQUARES), CP160/09, Universit\'{e} libre de Bruxelles, 1050 Brussels, Belgium}

\begin{abstract}
We report multiconfiguration Dirac-Hartree-Fock and relativistic configuration interaction calculations on the Thallium (Tl) electron affinity, as well as on the excited energy levels arising from the ground configuration of Tl$^-$.
The results are compared with the available experimental values and further validated by extending the
study to its homologous, lighter element, Indium (In), belonging to Group 13 (III.A) of the periodic table.
The calculated electron affinities of In and Tl, 383.4 and 322.8 meV, agree with the latest measurements by within 1\%.
Three bound states $^3P_{0,1,2}$ are confirmed in the $5s^25p^2$ configuration of In$^-$ while only the ground state $^3P_{0}$ is bound in the $6s^26p^2$ configuration of Tl$^-$.
The isotope shifts on the In and Tl electron affinities are also estimated.
The E2/M1 intraconfiguration radiative transition rates within $5s^25p^2 \; ^3P_{0,1,2}$ of In$^-$  are used to calculate the radiative lifetimes of the metastable $^3P_{1,2}$ levels.

\end{abstract}
\maketitle

\section{Introduction}
Negative ions play a major role in a number of areas of physics and chemistry involving ionized gases and plasma. Since there is no long-range Coulomb interaction between the outermost electron and the atomic core, their properties critically depend on electron-electron correlation and polarization and negative ions will only have a few bound states.
In most cases, the latter have the same parity or even belong to the same electronic configuration.
Only in a few cases, namely, Os$^-$~\cite{Bilodeau2000,Warring2009,Fischer2010}, Ce$^-$~\cite{Walter2007,Walter2011}, La$^-$~\cite{Walter2014,Jordan2015,Cerchiari2018}, and Th$^-$~\cite{Tang2019}, negative ions have excited bound states of opposite parity to that of the ground state making them good candidates for laser cooling. The most promising ones are so far La$^-$ and Th$^-$.

Several experimental techniques are possible to measure atomic electron affinities (EAs) and excited energy levels of negative ions with high precision.
Nevertheless some atomic electron affinities and anion fine-structure splittings are still known with limited accuracy.
All elements of Group 13 (B, Al, Ga, In, and Tl) form stable negative ions with electron affinities of less than 0.5 eV. The latter are therefore challenging to determine with accuracy, especially for the heavier elements.
Recently, tunable laser photodetachment threshold spectroscopy (LPTS) was used to measure the electron affinity of the $6s^26p^2\ ^3\!P_0$ ground state of $^{205}$Tl$^-$ to be 320.053(19)~meV~\cite{Walter2020}, which differs significantly from the value of  377(13)~meV obtained by a fixed-frequency laser photodetachment electron spectroscopy (LPES) measurement~\cite{Carpenter2000}.
Both experiments indicate that the excited levels are either unbound or too weakly bound
to be detected, although the
three fine-structure levels $^{3}P_{0,1,2}$ were detected as bound states for the lighter elements of the same Group 13 (B$^-$, Al$^-$, Ga$^-$ and In$^-$)~\cite{Williams1998,Scheer1998,Scheer1998a,Gibson2019,Tang2020,Walter2010}.
Number of theoretical studies on the EA of Tl have been reported using a variety of computational methods~\cite{Arnau1992,Wijesundera1997,Eliav1997,Guo-Xin1999,Figgen2008,Li2012,Felfli2012,Finney2019}, but their results show poor agreement.
For example, using different theoretical methods, Arnau et al.~\cite{Arnau1992} and  Felfli et al.~\cite{Felfli2012} predicted the EA of Tl to be 270 meV and 2415 meV, respectively.

In the present study, we resolve the disagreement between  experimental and theoretical EA values of Tl, and explore the existence of bound excited states of Tl$^-$.
Since the ground configurations of In ($5s^25p$) and In$^-$ ($5s^25p^2$) are analogous to those of Tl ($6s^26p$) and Tl$^-$ ($6s^26p^2$), we use the In/In$^-$ system  as a benchmark to support our  Tl/Tl$^-$ analysis.
We also estimate the balance between the nuclear mass and volume contributions to  the isotope shift (IS) on electron affinities of different isotopes of In and Tl.
The radiative lifetimes of the excited In$^-$ $5s^25p^2 \; ^3P_{1,2}$ fine structure levels based on the intraconfiguration radiative decay rates are reported.

\section{Theory}
\subsection{Multiconfiguration Dirac-Hartree-Fock approach}
The MCDHF method~\cite{Fischer2016}, as implemented in the {\sc Grasp} computer  package~\cite{Jonsson2013,Fischer2019}, is employed to obtain wave functions that are referred to as atomic state functions (ASFs), i.e., approximate eigenfunctions of the Dirac-Coulomb Hamiltonian given by
\begin{equation}
H_{\rm DC}= \sum_{i=1}^N(c~\bm{\alpha_i}\cdot\bm{p_i}+(\beta_i-1)c^2+V_i)+\sum_{i<j}^N\frac{1}{r_{ij}},
\end{equation}
where $V_i$ is the monopole part of the electron-nucleus interaction for a finite nucleus, $r_{ij}$ is the distance between electrons $i$ and $j$, $\bm{\alpha}$ and $\beta$ are the Dirac matrices.

Electron correlation  is included by expanding $\Psi\left(\Gamma PJ\right)$, an ASF, over a linear combination of configuration state functions (CSFs) $\Phi\left(\gamma_i PJ\right)$,
%COMMENT: just because you defined ASF using small letters ;-)
\begin{equation}
\Psi\left(\Gamma PJ\right)=\sum_{i=1}^Mc_i\Phi\left(\gamma_i PJ\right),
\end{equation}
where $\gamma_i$ represents all the coupling tree quantum numbers needed to uniquely define the CSF, besides the parity $P$ and the total angular momentum $J$. The CSFs are four-component spin-angular coupled, antisymmetric products of Dirac orbitals
of the form
\begin{equation}
\phi({\bf r})=\frac{1}{r}\left(\begin{array}{c}P_{n\kappa}(r)\chi_{\kappa m                            }(\theta,\phi)\\
iQ_{n\kappa}(r)\chi_{-\kappa m}(\theta,\phi)\end{array}\right).
\end{equation}
The radial parts of the one-electron orbitals and the expansion coefficients $c_i$ of the CSFs are obtained by the relativistic self-consistent field (RSCF) procedure.
In the present paper, the CSF expansions are obtained using the restricted active set (RAS) method, by allowing single and double (SD) substitutions from a selected set of reference configurations to a given orbital active set. The latter is systematically expanded by the addition of successive orbital layers to monitor the convergence of the calculated energies or any other relevant observable.

Each RSCF calculation is followed by a relativistic configuration interaction (RCI) calculation, where the Dirac orbitals are kept fixed and only the expansion coefficients of the CSFs are determined for selected eigenvalues and eigenvectors of the complete interaction matrix.
In this procedure, the Breit interaction and leading quantum electrodynamic (QED) effects (vacuum polarization and self-energy) are included.

In addition to energy levels, lifetimes $\tau$ and transition parameters, such as transition rates $A$ and line strengths $S$, are also computed. The transition parameters between two states $\gamma' P'J'$ and $\gamma PJ$ are expressed in terms of reduced matrix elements of the relevant transition operators~\cite{Gra:74a,Cowan1981}:
\begin{equation}
\label{eq:T_me}
\langle \Psi(\gamma PJ)||\textbf{T}||\Psi(\gamma' P'J') \rangle=\sum_{k,l}c_kc'_l \langle \Phi(\gamma_k PJ)||\textbf{T}||\Phi(\gamma'_l P'J') \rangle.
\end{equation}
Biorthogonal orbital transformations and CSF-expansion counter-transformations are used \cite{Olsetal:95a} when radial non-orthogonalities arise from the independent optimization of the two ASFs involved in (\ref{eq:T_me}).

\subsection{Isotope shift}
We  define the isotope shift (IS) on the EA between two isotopes of mass $A$ and $A'$ as
\begin{equation}
{\rm IS(EA)}^{AA¡¯} ={\rm EA}(A)-{\rm EA}(A') \; ,
\end{equation}
in agreement with the frequency isotope shift definition adopted in the description of the RIS3~\cite{Naze2013} and RIS4~\cite{Ekman2019} codes.
In the $A>A'$ case, a positive isotope shift on the EA implies a larger electron affinity for the heavier isotope. Such an IS is qualified as a ``normal'' IS, referring to the normal mass shift encountered for the one-electron atomic hydrogen characterized by a blueshift %(towards smaller wavelengths)
of the spectral lines of deuterium $^2$H compared with hydrogen $^1$H.
The IS can be decomposed into two contributions: the mass shift (MS) and the field shift (FS). They arise, respectively, from the recoil effect due to the finite mass of the nucleus, and from the difference in nuclear charge distributions between the two isotopes.
The revised version of the RIS3 code~\cite{Naze2013}, RIS4, based on a reformulation of the field shift~\cite{Ekman2019}, is used for the computations of IS parameters in the present paper.

\subsubsection{Mass shift}
The isotope mass shift of an atomic level $i$ is obtained by evaluating the expectation values of the
operator,
\begin{equation}
\label{eq:H_MS}
H_{\rm MS}=\frac{1}{2M}\sum_{j,k}^{N}\left( \bm{p_j \cdot p_k}
-\frac{\alpha Z}{r_j}\left(\bm{\alpha_j} + \frac{(\bm{\alpha_j \cdot r_j})\bm{r_j}}{r_j^2} \right)  \cdot \bm{p_k} \right),
\end{equation}
where $M$ is the nuclear mass of the isotope~\footnote{Nuclear masses are calculated by subtracting the mass of the electrons from the atomic masses \cite{Audi2003} and  adding the total binding electronic energy using the prescriptions of \cite{Filetal:2016b}}.

Separating the operator into one-body and two-body terms, $H\rm{_{MS}}$ can be rewritten as the  sum of normal mass shift (NMS) and specific mass shift (SMS) contributions,
\begin{equation}
H_{\rm MS}=H_{\rm NMS}+H_{\rm SMS} \ .
\end{equation}

The (mass-independent) normal mass shift $K\rm{_{NMS}}$ and specific mass shift $K\rm{_{SMS}}$ parameters for a level $i$ are defined by the following expressions
\begin{equation}
\frac{K_{\rm \emph{i},NMS}}{M}\equiv \langle \Psi_i(\gamma PJ)|H_{\rm NMS}|\Psi_i(\gamma PJ) \rangle \ ,
\end{equation}
\begin{equation}
\frac{K_{\rm \emph{i},SMS}}{M}\equiv\langle \Psi_i(\gamma PJ)|H_{\rm SMS}|\Psi_i(\gamma PJ) \rangle\ .
\end{equation}
%By analogy with (\ref{eq_NMS}) and (\ref{eq_SMS}),
The mass shift parameters can be decomposed into three parts,
\begin{equation}
K_{\rm \emph{i},NMS}=K^1_{\rm \emph{i},NMS}+K^2_{\rm \emph{i},NMS}+K^3_{\rm \emph{i},NMS}\ ,
\end{equation}
\begin{equation}
K_{\rm \emph{i},SMS}=K^1_{\rm \emph{i},SMS}+K^2_{\rm \emph{i},SMS}+K^3_{\rm \emph{i},SMS}\ .
\end{equation}
where the $K^1$ terms refer to the ``uncorrected" relativistic contributions (first term of~(\ref{eq:H_MS})) and the sum $K^2+K^3$ (2nd and 3rd terms of (\ref{eq:H_MS})) to the lowest-order relativistic corrections in the Breit approximation~\cite{Sha:85a,ShaArt:94a}.
The mass shift contribution to the IS on the EA
\begin{equation}
\label{eq:IS_EA_MS}
    {\rm{IS}(EA)}^{AA¡¯}_{\rm{MS}} =
     \Delta K_{\rm MS}
     \left( \frac{1}{M} - \frac{1}{M'} \right) \; ,
\end{equation}
is therefore directly proportional to the difference of the MS electronic parameters
\begin{equation}
\label{eq:Deta_K_MS}
    \Delta K_{\rm MS} = K_{\rm g.s., \ MS} - K^{-}_{\rm g.s., \ MS} \; ,
\end{equation}
where the minus exponent refers to quantities related to the negative ion and g.s. stands for ground state.

\subsubsection{Field shift}
In the first order perturbation approximation, the field shift for a given level $i$ can be expressed as
\begin{equation}
\label{eq_FS1}
\delta E_{i,\rm{FS}}^{(1)A,A'}=-\int_{\bm{R^3}} \left[V^A(\bm{r})-V^{A'}(\bm{r})\right ]\rho_i^e(\bm{r})d^3\bm{r} \ ,
\end{equation}
where $V^A(\bm{r})$ and $V^{A'}(\bm{r})$ are the potentials arising from the nuclear charge distributions of the two isotopes and $\rho_i^e(\bm{r})$ is the electron density.
By approximating the electron density at the origin with a spherically symmetric even polynomial function
\begin{equation}
\rho_i^e(\bm{r})\approx b_{i,0}+b_{i,2}r^2+b_{i,4}r^4+b_{i,6}r^6 \ ,
\end{equation}
Eq.~(\ref{eq_FS1}) can be expressed as
\begin{equation}\label{eq_FS2}
\delta E_{i,\rm{FS}}^{(1)A,A'}\approx \sum_{0\leq n\leq6,even} F_{i,n} \; \delta\langle r^{n+2}\rangle^{A,A'}
\end{equation}
where $F_{i,n}$ are level electronic factors and $\delta\langle r^{n}\rangle^{A,A'}=\langle r^{n}\rangle^{A}-\langle r^{n}\rangle^{A'}$.
Assuming a constant electron density within the nuclear volume, we get from the first term of Eq.~(\ref{eq_FS2})
\begin{equation}\label{eq_FS3}
\delta E_{i,\rm{FS}}^{(1)A,A'}\approx F_{i,0} \; \delta\langle r^{2}\rangle^{A,A'}.
\end{equation}

As suggested in Ref~\cite{Ekman2019},
we can include the effect of a varying electronic density (ved) to evaluate the FS, by introducing the appropriate corrected level electronic factors, $F_{i,0}^{(0)\rm{ved}}$ and $F_{i,0}^{(1)\rm{ved}}$, without considering higher order nuclear moments. Eq.~(\ref{eq_FS2}) is then replaced by
\begin{equation}\label{eq_FS4}
\delta E_{i,\rm{FS}}^{(1)A,A'}\approx \left(F_{i,0}^{(0)\rm{ved}}+F_{i,0}^{(1)\rm{ved}}\delta\langle r^{2}\rangle^{A,A'}\right)\delta\langle r^{2}\rangle^{A,A'} \; .
\end{equation}
The FS contribution to the  IS on the EA can therefore be estimated from
\begin{equation}
\label{eq:IS_EA_FS}
    {\rm{IS}(EA)}^{AA¡¯}_{\rm{FS}} = \left(F_{\rm{g.s.}, 0}-F_{\rm{g.s.}, 0}^{-}\right)\delta\langle r^{2}\rangle^{A,A'}
%= \Delta F_{0} \;  \delta\langle r^{2}\rangle^{A,A'} \; ,
\end{equation}
with
\begin{equation}
\label{eq:IS_EA_FS_ced}
\left(F_{\rm{g.s.}, 0}-F_{\rm{g.s.}, 0}^{-}\right) = \Delta F_{0}
\end{equation}
in the constant electron density approximation, or
\begin{equation}
\label{eq:IS_EA_FS_ved}
\left(F_{\rm{g.s.}, 0}-F_{\rm{g.s.}, 0}^{-}\right) =
\left( \Delta F_{0}^{(0)\rm{ved}} + \Delta F_{0}^{(1)\rm{ved}}\delta\langle r^{2}\rangle^{A,A'}\right)
\end{equation}
using the varying electronic density model.

\section{Results and discussions}
\subsection{Electron affinities}

To evaluate the ground state energies of the Tl neutral atom ([Xe]$4f^{14}5d^{10}6s^26p\ ^2P^{o}_{1/2}$) and Tl$^-$ anion ([Xe]$4f^{14} 5d^{10}6s^26p^2\ ^3P_0$), we start from single-reference (SR) calculations, where CSFs lists are generated by allowing single and double (SD) excitations from the $5d$, $6s$ and $6p$ electrons
to orbitals with $n\leq10$, $l\leq5$. The CSFs that contribute by more than 0.1\% in weight ($w_i = \vert c_i \vert^2$) to the ground states wave functions of Tl and Tl$^-$ are reported in Table~\ref{tab_c}.
These configurations
\begin{equation}
    \begin{aligned}
    \mathrm{Tl~(odd): }&\ 5d^{10} 6s^2 6p, \ 5d^8 5f^2 6s^2 6p, \ 5d^{10} 6p^3,
    \\
    \mathrm{Tl}^- \mathrm{~(even): }& \ 5d^{10} 6s^2 6p^2, \ 5d^8 5f^2 6s^2 6p^2, \ 5d^{10} 6p^4, \ 5d^{10} 6s 6p^2 6d, \ 5d^9 5f 6s 6p^2 7p
    \end{aligned}
\end{equation}
form the multireference spaces used in the following.

In the MR calculations, SD excitations are allowed from all MR configurations to an increasing active set (AS) of orbitals. The calculations are performed layer by layer, introducing at each step at most one new correlation orbital per angular {$\kappa$}-symmetry. Excitations from the  SR configuration are ultimately allowed to an active set of orbitals with $n\leq 13$ and $l \leq 5$, noted $13h$. Excitations from the remaining MR configurations are, however, limited to smaller active sets, i.e., $6s6p6d5f$ for the neutral atom and $8s8p7d6f$ for the anion, to keep the number of CSFs manageable. The slightly larger active set for the anion is required to balance correlation effects between the atom and its anion.
Two sets of calculated EAs with increasing ASs  are listed in Table~\ref{tab_EA}. One is EA$(\Delta n$=0)=$E_n$(Tl)$-E_n$(Tl$^-$), where $E_n$ labels the energy obtained with the AS of maximum principal quantum number $n$, the other is EA$(\Delta $=1)=$E_n$(Tl)$-E_{n+1}$(Tl$^-$), that  can be justified by the fact that more orbitals are needed to describe electron correlation for the negative ion than for the neutral atom. The former is increasing with active sets and provides a lower bound to the calculated EA, whereas the latter is decreasing and hence provides an upper bound.
We used a non-linear exponential decay function to extrapolate the last four EA$(\Delta n$=0) and EA$(\Delta n$=1) values, and adopted the intersection as our final theoretical EA value~\cite{Si2018}.
The theoretical uncertainty of the {\it ab initio} EA inevitably depends on the correlation models used for tailoring the ASF expansions. Based on our passed experience on complex systems \cite{Si2018,Tang2019} and on the comparison with observation~\cite{Tang2019,Leietal:2020a}, we estimated it to be less than 2\%. This conservative estimation covers the 0.8\% corresponding to half of the interval between the two EA$(\Delta n = 0,1)$ values and the smaller uncertainty (0.65\%) associated with the extrapolation procedure.
With this uncertainty estimation, our final thallium EA-value is 322.8(6.5)~meV.

A comparison between experimental and theoretical EA values is presented in  Table~\ref{tab_EA2}. One can see
that our final thallium EA-value of 322.8(6.5)~meV agrees with the very recent LPTS experimental value of 320.053(19)~meV~\cite{Walter2020}, but definitely lies outside the error bars of the previous LPES measurement 377(13)~meV~\cite{Carpenter2000}.
As mentioned in the introduction, the scattering of theoretical results is surprisingly large. Amongst the most recent works, our MCDHF-RCI value is in good agreement with the results of Finneyet {\it et al.}~\cite{Finney2019} using the relativistic coupled-cluster version of the Feller-Peterson-Dixon composite method  (RCC-FPD). Our theoretical estimation is definitely smaller -~by almost one order of magnitude~- than the complex angular momentum (CAM) electron elastic total cross-sections result of Felfli {\it et al.}~\cite{Felfli2012}. The last authors  suggested that all the  EA values of thallium reported in the literature before their work should be considered as the binding energy of an excited state of the anion. All RCI calculations performed in the present work for the odd parity exclude this possibility, the lowest state, $5d^{10} 6s 6p^3 \; ^5S_2$, being estimated to lie around 5.2~eV above the ground level of Tl$^-$.

Since In and In$^-$~$(Z=49)$ are homologous elements to Tl and Tl$^-$~$(Z=81)$,  we performed similar calculations for the electron affinity of In to further validate our calculated EA of thallium.
SD excitations from $4d^{10}5s^25p$ and $4d^{10}5s^25p^2$ are allowed up to orbitals with $n\leq12$, $l\leq5$; SD excitations from $4d^{8}4f^25s^25p$ and $4d^{10}5p^3$ are allowed up to the $5s5p5d4f$ AS for the In neutral atom;  SD excitations from $4d^{8}4f^25s^25p^2$, $4d^{10}5p^4$, $4d^{10}5s5p^25d$ and $4d^{9}5f^25s5p^26p$ are included up to the $7s7p6d5f$ AS for the In$^-$ anion.
The calculated EA$(\Delta n$=0) and EA$(\Delta n$=1) values are also listed in Table~\ref{tab_EA}.  Using the same extrapolation method as for Tl$^-$, we obtain a final value of 383.(7.7)~meV for which we adopted the 2\% uncertainty estimation, as discussed above, that largely covers  half of the interval between the two EA$(\Delta n = 0,1)$ values (1.1\%). This theoretical result is also in excellent agreement with the LPTS experimental value of 383.92(6)~meV~\cite{Walter2010}. Similarly as for thallium, our indium EA value lies outside the confidence interval of the LPES measurement of 404(9)~meV~\cite{Williams1998}.

\begin{table}
\setlength{\tabcolsep}{3pt}
\centering
\caption{Wave function compositions for the ground states of Tl and Tl$^-$. See text for the definition of the weights $w_i$.}\label{tab_c}
\begin{tabular}{cccccccccccccc}
\hline
\hline
\multicolumn{3}{c}{Tl} & \multicolumn{3}{c}{Tl$^-$}\\
\cmidrule(lr){1-3} \cmidrule(rl){4-6}
Configuration & Term &$w_i$(\%) & Configuration & Term & $w_i$(\%)  \\
\hline
    $5d^{10}6s^26p$                  & $^2P^{o}$ & 92.1 &
$5d^{10}6s^26p^2$                & $^3P$ & 81.9\\

$5d^{8}(^3P)5f^{2}(^1S)6s^26p$   & $^2P^{o}$ & 0.61 &
$5d^{10}6s^26p^2$                & $^1S$ & 8.84 \\

$5d^{8}(^3F)5f^{2}(^1S)6s^26p$   & $^2P^{o}$ & 0.53 &
$5d^{8}(^3P)5f^{2}(^1S)6s^26p^2$ & $^3P$ & 0.55\\

$5d^{10}6p^3$                    & $^2P^{o}$ & 0.47 &
$5d^{8}(^3F)5f^{2}(^1S)6s^26p^2$ & $^3P$ & 0.48\\

$5d^{8}(^1D)5f^{2}(^1S)6s^26p$   & $^2P^{o}$ & 0.25 &
$5d^{8}(^1D)5f^{2}(^1S)6s^26p^2$ & $^3P$ & 0.22\\

$5d^{8}(^1S)5f^{2}(^1S)6s^26p$   & $^2P^{o}$ & 0.15 &
$5d^{10}6p^4$                    & $^3P$ & 0.18\\

$5d^{8}(^1G)5f^{2}(^1S)6s^26p$   & $^2P^{o}$ & 0.14 &
$5d^{8}(^1S)5f^{2}(^1S)6s^26p^2$ & $^3P$ & 0.13\\

  & & & $5d^{8}(^1G)5f^{2}(^1S)6s^26p^2$ & $^3P$ & 0.12\\
  & & & $5d^{10}6s6p^2(^2D)6d$ & $^3P$ & 0.12\\
  & & & $5d^{9}5f(^1P)6s(^2P^{o})6p^2(^4D)7p$ & $^3P$ & 0.11\\
  & & & $5d^{9}5f(^1P)6s(^2P^{o})6p^2(^2D)7p$ & $^3P$ & 0.10\\
  & & & $5d^{9}5f(^1P)6s(^2P^{o})6p^2(^4P)7p$ & $^3P$ & 0.10\\
\hline
\end{tabular}
\end{table}

\begin{table}
\setlength{\tabcolsep}{4pt}
\footnotesize
\centering
\caption{Theoretical electron affinities of In and Tl (present work). All values in meV.}\label{tab_EA}
\begin{tabular}{ccccccccc}
\hline
\hline
\multicolumn{3}{c}{In} & \multicolumn{3}{c}{Tl} \\
 \cmidrule(lr){1-3} \cmidrule(lr){4-6}
 AS & EA($\Delta n$=0) & EA($\Delta n$=1) & AS & EA($\Delta n$=0) & EA($\Delta n$=1)  \\
\hline
9h  & 356.2 & 428.6 & 10h & 303.3 & 342.1\\
10h & 368.5 & 398.6 & 11h & 311.9 & 327.9\\
11h & 374.4 & 390.0 & 12h & 315.9 & 324.7\\
12h & 378.1 & 386.5 & 13h & 318.5 & 323.7\\
Final&\multicolumn{2}{c}{383.4(7.7)} & & \multicolumn{2}{c}{322.8(6.5)}\\
\hline
\end{tabular}
\end{table}

\begin{table}
\setlength{\tabcolsep}{4pt}
\footnotesize
\centering
\caption{Comparison of the present calculated electron affinities of In and Tl with the experimental results and other theoretical values. LPTS: laser photoelectron threshold spectrocopy, LPES: laser photodetachment electron spectroscopy, MCDHF: Multiconfiguration Dirac-Hartree-Fock calculations (present work), CIPSI: multireference single and double configuration-interaction method, RCC: relativistic coupled cluster, HFR-DFT: pseudo-relativistic Hartree¨CFock and density functional theory, IHFSCC: intermediate-Hamiltonian Fock-space coupled cluster method, CAM: complex angular momentum method,  RCC-FPD: relativistic coupled-cluster version of the Feller-Peterson-Dixon composite method. (All values in meV.) \label{tab_EA2}}
\begin{tabular}{ccccccccc}
\hline
\hline
  & Method &  EA(In) & EA(Tl) \\
\hline
\hline
\multicolumn{4}{c}{Experiment}  \\
\hline
Walter et al.~\cite{Walter2010} & LPTS & 383.92(6)\\
Walter et al.~\cite{Walter2020} & LPTS & & 320.053(19)\\
Williams et al.~\cite{Williams1998} & LPES & 404(9)\\
Carpenter et al.~\cite{Carpenter2000} & LPES & & 377(13) \\
%& & & \\
\hline
\multicolumn{4}{c}{Theory}  \\
\hline
 Present work  & MCDHF  & 383.4(7.7) & 322.8(6.5) \\
Wijesundera~\cite{Wijesundera1997}  & MCDHF  & 393 & 291 \\
Li et al.~\cite{Li2012}             & MCDHF  & 397.83 & 290.20 \\
Arnau et al.~\cite{Arnau1992}       & CIPSI  & 380 & 270 \\
Eliav et al.~\cite{Eliav1997}       & RCC    & 419 & 400 \\
Chen and Ong~\cite{Guo-Xin1999}     & HFR-DFT& 429 & 388 \\
Figgen et al.~\cite{Figgen2008}     & IHFSCC & 403 & 347 \\
Felfli et al.~\cite{Felfli2012}     & CAM  & 380 & 2415 \\
Finneyet al.~\cite{Finney2019}      & RCC-FPD & 386 & 320\\
\hline
\end{tabular}
\end{table}
%\clearpage
\subsection{Isotope shifts on the Indium and Thallium  electron affinities}\label{sec_EA}
Both In and Tl have many isotopes. While the stable isotope $^{113}$In is only 4.3\% of naturally occuring indium, in lower abundance than the long-lived radioactive isotopes, Tl has two stable isotopes, $^{203}$Tl (30\% natural abundance) and $^{205}$Tl (70\%). In this section, we report the isotope shifts on the electron affinity of In and Tl by using the wave functions obtained for estimating the electron affinity (see previous section).
 The differences of the isotope shift mass and field electronic parameters that make the IS on the EA (see Eqs. (\ref{eq:Deta_K_MS} and (\ref{eq:IS_EA_FS}))   are listed in Table~\ref{tab_IS1}.

From this table, we can see  that for In, the sum of the relativistic corrections to the uncorrected one-electron normal mass shift operator, $\Delta(K^2_{\rm NMS}+K^3_{\rm NMS})$, reinforces  the $\Delta K^1_{\rm NMS}$ value by around $50\%$, while for the SMS, $\Delta(K^2_{\rm SMS}+K^3_{\rm SMS})$ counterbalances  $\Delta K^1_{\rm SMS}$ by $85\%$, leading to a large dominance of $\Delta K_{\rm NMS}$ over $\Delta K_{\rm SMS}$.

Relativistic corrections to the recoil operator play an even more important role in Tl, for which  $\Delta(K^2_{\rm NMS}+K^3_{\rm NMS})$ is 2.6 times larger than the uncorrected $\Delta K^1_{\rm NMS}$ value and strongly strengthen it.
For the SMS contribution, the $\Delta(K^2_{\rm SMS}+K^3_{\rm SMS})$  is $3.6$ times larger than $\Delta K^1_{\rm SMS}$ but of opposite sign.
Oppositely to In, the total $\Delta K_{\rm SMS}$ value is large, 66\% of the $\Delta K_{\rm NMS}$, and the constructive addition of both contributions makes the total  $\Delta K_{\rm MS}$ value $1.66$ times larger than the NMS contribution. \\

The differences of the electronic FS parameters are reported in the same Table~\ref{tab_IS1} for both In/In$^-$ and Tl/Tl$^-$ systems. Positive $\Delta F$ values reveal a  gain in electron density at the nucleus when detaching the outer electron from the anion. One should observe  that this gain factor is 10 times larger for thallium than for indium. The ratio of the Tl and In nuclear charges $ Z({\rm Tl})/Z({\rm In})
%= 81/49
\approx 1.65$, arising from the explicit linear $Z$-dependence of the FS factor~\cite{Naze2013,Ekman2019}, can only explain a little portion of this difference. The much larger remaining part of this factor ten is simply due to the Tl-In difference of the electron density within the nuclear volume. The FS contribution to the IS on the EA has been estimated from Eqs.(\ref{eq:IS_EA_FS}) and
(\ref{eq:IS_EA_FS_ved}) to include the effect of a varying electronic density, using the root mean square (rms) nuclear radii of from Ref.~\cite{Angeli2013}. \\

\begin{figure}
  \centering
  \includegraphics[width=\textwidth]{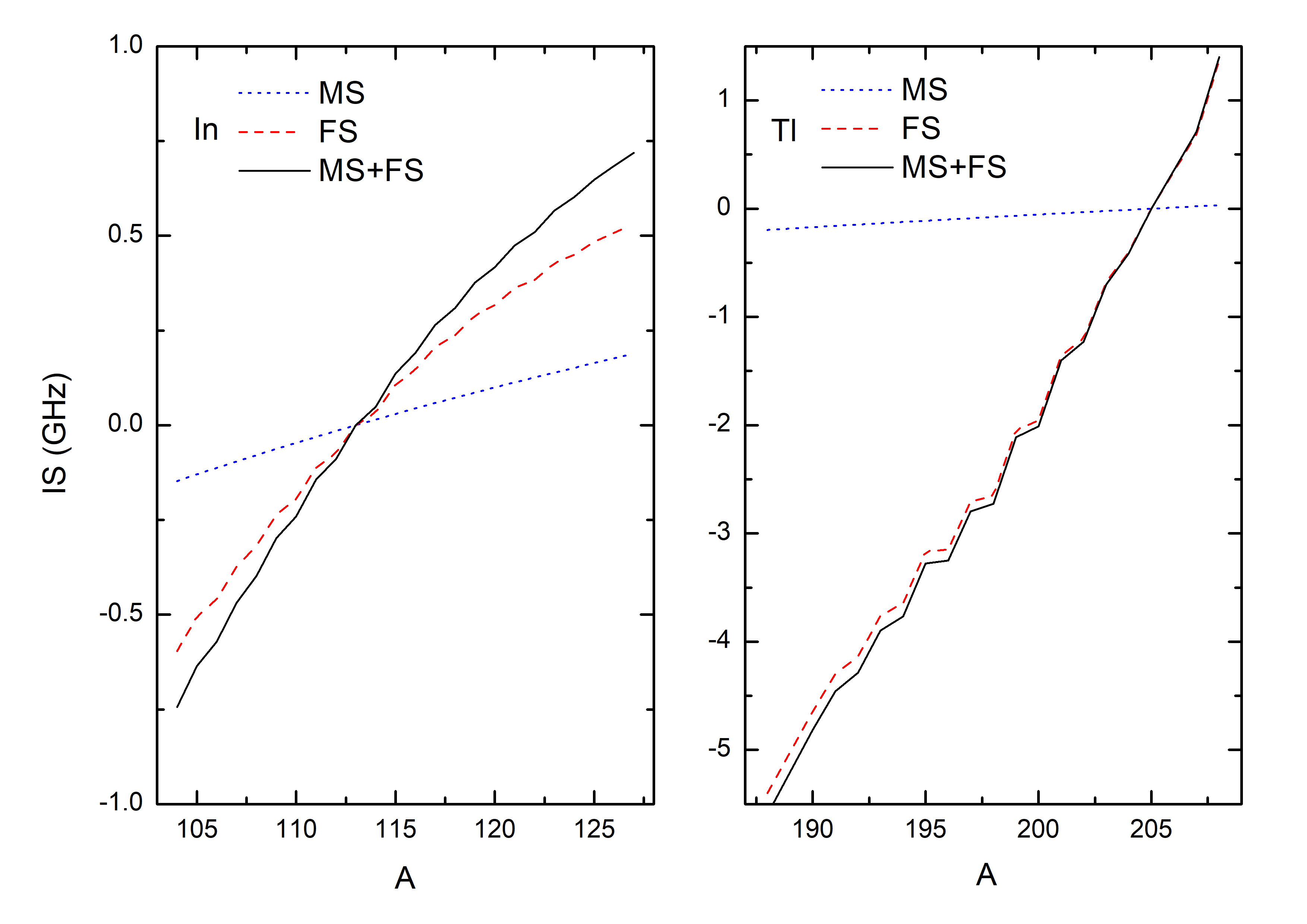}
  \caption{Mass (MS), Field (FS) and total (MS+FS) isotope shifts (IS)  on EAs relative to $^{113}$In and $^{205}$Th, i.e., EA($^{A}$In) - EA($^{113}$In) and EA($^{A}$Tl) - EA($^{205}$Th)). }
\label{fig:IS_versus_A}
\end{figure}

\clearpage
The mass, field and total isotope shifts on electron affinities are reported in Table~\ref{tab_IS2} and displayed in Fig.~\ref{fig:IS_versus_A}  for a large range of isotopes relative to the  $^{113}$In and $^{205}$Tl stable isotopes.

We observe that for EA(In), the FS is already more important than the MS, while for EA(Tl), the FS largely dominates the MS that becomes almost negligible. This is expected for heavy elements, as the mass factor $1/M-1/M'=(M'-M)/MM'$ decreases rapidly with the nuclear mass. For example, the MS contribution to the electron affinities  between the two stable isotopes of Tl, $^{203}$Tl and $^{205}$Tl, is ${\rm IS(EA)}^{205,203}_{\rm MS}=0.0214$~GHz.
If we assume a constant electron density within the nuclear volume, Eq.(\ref{eq_FS3}) gives a FS of
${\rm IS(EA)}^{205,203}_{\rm FS}=0.7334$~GHz. By including the effect of a varying electronic density, Eq.(\ref{eq_FS4}) gives a FS of 0.6799~GHz, which is 7\% lower than the former value. This observation is consistent with previous studies on IS (see e.g.,~\cite{Ekman2019, Schiffmann2021}) and emphasizes the need to include this effect. Adding the MS and latter FS contributions together, we  obtain that EA($^{205}$Tl) is 0.7014~GHz higher than EA($^{203}$Tl), i.e. ${\rm IS(EA)}^{205,203}=0.7013$~GHz.
\begin{table}
\setlength{\tabcolsep}{4pt}
\small
\centering
\caption{Isotope shift parameters on the electron affinities of In and Tl. $\Delta K$ values in GHz*u. $\Delta F_0$ and
$\Delta F_0^{ved}$ in GHz/fm$^2$, $\Delta F_0^{(1)ved} $ in GHz/fm$^4$.}\label{tab_IS1}
\begin{tabular}{lcccccccccccc}
\hline
\hline
 &  In/In$^-$
 &  Tl/Tl$^-$ \\
\hline
$\Delta K^1_{\rm NMS}$                     & -134   &  -74\\
$\Delta ( K^2_{\rm NMS}+K^3_{\rm NMS} ) $           &  -65   & -194\\
$\Delta K_{\rm NMS} $ & -199   & -268\\
%\hline
& & \\
$\Delta K^1_{\rm SMS}$                     &  41  &  68\\
$\Delta  ( K^2_{\rm SMS}+K^3_{\rm SMS} ) $           & -34  & -245\\
$\Delta K_{\rm SMS} $ &   6  & -178 \\
& & \\
$\Delta K_{\rm MS} = \Delta K_{\rm NMS} + \Delta K_{\rm SMS} $                    &  -192  &  -445 \\
\hline
$\Delta F_0$                           & 0.814   &  7.21\\
$\Delta F_0^{(0)ved}$                     & 0.791   &  6.68\\
$\Delta F_{0}^{(1)ved}$

& 4.47E$-$4 &  7.11E$-$3\\
\hline
\end{tabular}
\end{table}

\begin{table}
\setlength{\tabcolsep}{4pt}
\small
\centering
\caption{Mass (MS), Field (FS) and total (MS+FS) isotope shifts  on EAs relative to the most abundant isotopes, i.e, EA($^{A}$In) - EA($^{113}$In) and EA($^{A}$Tl) - EA($^{205}$Th). All shifts in GHz.} \label{tab_IS2}
\begin{tabular}{ccccccccccccc}
\hline
\hline
\multicolumn{4}{c}{In} & \multicolumn{4}{c}{Tl} \\
 \cmidrule(lr){1-4} \cmidrule(lr){5-8}
A & MS  & FS  & MS+FS & A & MS  & FS  & MS+FS\\
\hline
104 & -0.147 & -0.596 & -0.744 & 188 & -0.197 & -5.397 & -5.594 \\
105 & -0.130 & -0.505 & -0.635 & 190 & -0.172 & -4.645 & -4.816 \\
106 & -0.113 & -0.459 & -0.572 & 191 & -0.159 & -4.297 & -4.456 \\
107 & -0.096 & -0.374 & -0.469 & 192 & -0.147 & -4.137 & -4.285 \\
108 & -0.079 & -0.318 & -0.397 & 193 & -0.135 & -3.76  & -3.895 \\
109 & -0.063 & -0.236 & -0.298 & 194 & -0.123 & -3.644 & -3.767 \\
110 & -0.047 & -0.195 & -0.241 & 195 & -0.112 & -3.165 & -3.276 \\
111 & -0.031 & -0.112 & -0.143 & 196 & -0.1   & -3.15  & -3.250 \\
112 & -0.015 & -0.075 & -0.090 & 197 & -0.088 & -2.707 & -2.795 \\
{\bf 113} &  {\bf 0.0}      &     {\bf 0.0}      &   {\bf 0.0}  & 198 & -0.077 & -2.649 & -2.725 \\
114 & 0.015  & 0.034  & 0.048  & 199 & -0.066 & -2.044 & -2.110 \\
115 & 0.03   & 0.106  & 0.136  & 200 & -0.054 & -1.957 & -2.011 \\
116 & 0.044  & 0.147  & 0.191  & 201 & -0.043 & -1.359 & -1.402 \\
117 & 0.058  & 0.206  & 0.264  & 202 & -0.032 & -1.198 & -1.231 \\
118 & 0.072  & 0.237  & 0.310  & 203 & -0.021 & -0.680 & -0.701 \\
119 & 0.086  & 0.290  & 0.376  & 204 & -0.011 & -0.402 & -0.413 \\
120 & 0.100  & 0.317  & 0.416  & {\bf 205} &  {\bf 0.0}      & {\bf 0.0}     &  {\bf 0.0}     \\
121 & 0.113  & 0.362  & 0.475  & 207 & 0.021  & 0.688  & 0.709  \\
122 & 0.126  & 0.384  & 0.510  & 208 & 0.031  & 1.370  & 1.402  \\
123 & 0.139  & 0.428  & 0.567  &     &        &        &        \\
124 & 0.152  & 0.451  & 0.602  &     &        &        &        \\
125 & 0.164  & 0.484  & 0.648  &     &        &        &        \\
126 & 0.176  & 0.507  & 0.684  &     &        &        &        \\
127 & 0.188  & 0.530  & 0.719  &     &        &        &        \\
\hline
\end{tabular}
\end{table}

\subsection{$ns^2np^2$ levels of In$^-~(n=5)$ and  Tl$^-~(n=6)$}

\begin{figure}
  \centering
  \includegraphics[width=\textwidth]{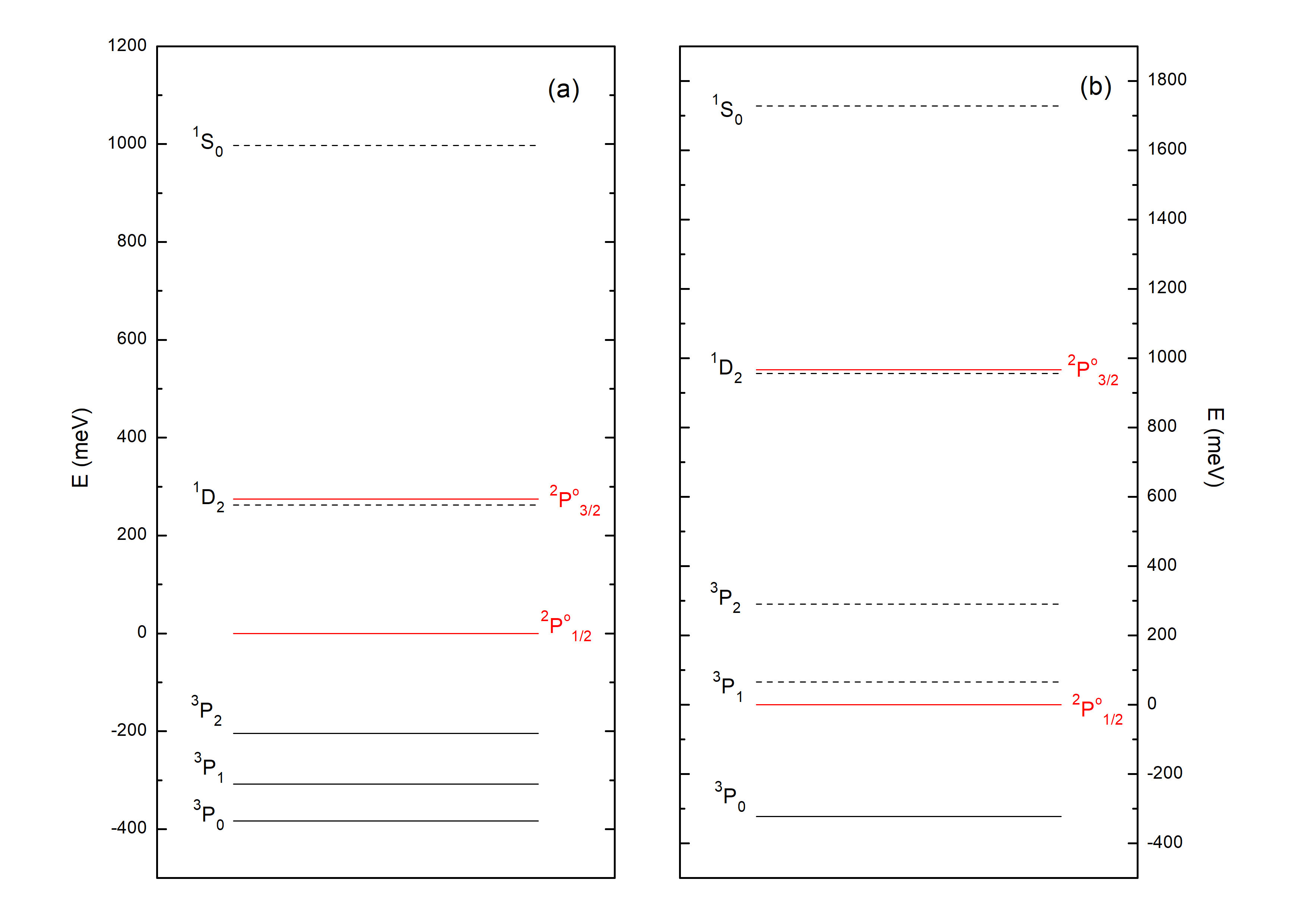}\\
  \caption{Energy diagram of $5s^25p^2$ levels of In$^-$ (a) and $6s^26p^2$ levels of Tl$^-$ (b) relative to the ground states of In and Tl, respectively. The fine-structure splitting of $^2P^{o}_{1/2,3/2}$ for the neutral atoms are from NIST ASD~\cite{NIST_ASD}.}\label{fig_E}
\end{figure}

The RCI excitation energies of the four levels $5s^25p^2 \; ^3P_{1,2}, \; ^1D_2$ and $^1S_0$ of In$^-$ based on the wave functions described in section~\ref{sec_EA}, are reported in Table~\ref{tab_E}.
The In$^-$ energy levels relative to the ground state of In ($5s^25p\ ^2P^{o}_{1/2}$) are also displayed in Fig.~\ref{fig_E}(a).

The excited energy levels excited levels, $5s^2 2p^2 \; ^3P_{1}$ and $^3P_{2}$ of In$^-$ have been observed as being stable using the techniques of laser-photo\-detachment electron spectroscopy (LPES)~\cite{Williams1998} and laser-photodetachment threshold spectroscopy (LPTS)~\cite{Walter2010}.
Our theoretical work confirms the existence of three bound states in In$^-$, all belonging to the $^3P$ fine structure. The theory-observation agreement with observation is quite satisfactory.
The presently calculated $^3P_1$ energy level agrees within 0.5~meV with the two experimental values while the $^3P_2$ energy level is predicted to be 9~meV higher than the result of the LPTS measurement~\cite{Williams1998}.

Similar RCI calculations were performed for the  energy levels $6s^26p^2 \; ^3P_{1,2}, \; ^1D_2$ and $^1S_0$ of Tl$^-$. The corresponding excitation energies  are  displayed relatively to the Tl ground state ($6s^26p\ ^2P^{o}_{1/2}$) in Fig.~\ref{fig_E}(b).
We can see that due to the large fine-structure splitting of $6s^26p^2\ ^3P_{0,1,2}$, the $^3P_{0}$ level is the only bound state in Tl$^-$, which agrees with the interpretation of the recent threshold spectroscopy  measurements~\cite{Walter2020}.

The $^1D_{2}$ and $^1S_{0}$ levels are both unbound in In$^-$ and Tl$^-$ (see Fig.~\ref{fig_E}). For both systems, the $^1D_{2}$ level of the anion and  the $^2P^{o}_{3/2}$ level of the corresponding neutral atom are almost degenerate. \\

Recent progress has been done in the measurements of radiative lifetimes of metastable levels of negative ions using cold storage techniques \cite{Bacetal:2015a,Zet:2017a}. It is therefore
worthwhile to report the theoretical lifetimes of the  In$^-$~$5s^25p^2\ ^3P_{1,2}$ levels that have not been measured yet so far. Our predicted lifetimes, based on our theoretical M1 and E2 radiative decay rates are reported in Table~\ref{tab_E}. Looking at the selection rules~\cite{Cowan1981}, the $^3P_{1}$ level can only decay to the ground state $^3P_{0}$ through a magnetic dipole (M1) process. The   $^3P_{2}$ level can decay to $^3P_0$ via an electric quadrupole (E2) radiative transition, and to $^3P_{1}$ through both M1/E2 de-excitations but the E2 transition  probabilities  are found to be much smaller than the M1 amplitudes  by at least 2-3 orders of magnitude. This means that the theoretical lifetimes  mostly depend on the M1 rates that usually quickly converge with the correlation models~\cite{Suetal:2019a}. The  quality of the transition energies is however an important ingredient due to the $\lambda^3$ scaling factor appearing in the M1  spontaneous emission A rate. For this reason, we also report the adjusted lifetimes to the experimental excitation energies~\cite{BraGru:2016a}, measured in the  LPTS~\cite{Walter2010} experiment.
The resulting lifetimes of a few hundreds of seconds, could be measured in a cryogenic ion storage ring that was demonstrated to be efficient to store negative ion beams in the hour time domain~\cite{Bacetal:2015a,Zet:2017a}.

\begin{table}
\setlength{\tabcolsep}{3pt}
\footnotesize
\centering
\caption{Excitation energies (in meV) and radiative lifetimes ($\tau$, in s) of the  In$^-$ $5s^25p^2\; ^3P_{1,2}$ levels. MCDHF: present work, LPTS: laser photoelectron threshold spectroscopy measurements~\cite{Walter2010}, LPES: laser photodetachment electron spectroscopy~\cite{Williams1998},
MCDHF (adj.): adjusted lifetimes using the LPTS experimental transition energies~\cite{Walter2010}.}
\label{tab_E}
\begin{tabular}{ccccccccccc}
\hline
\hline
& \multicolumn{3}{c}{Excitation energies (meV)} && \multicolumn{2}{c}{$\tau$ (s)} \\
 \cmidrule(lr){2-4} \cmidrule(lr){6-7}
 & MCDHF & LPTS & LPES && MCDHF & MCDHF (adj.)  \\
\hline
$^3P_1$ & 75.43 & 76.06(7) & 76(9) && 251.8 & 245.6 &  \\
$^3P_2$ & 179.4 & 170.6(6) & 175(9)&~~~& 129.8 & 172.7 &  \\
\hline
\end{tabular}
\end{table}

\clearpage
\section{Conclusion}
In summary, we calculated the EAs of In and Tl to be 383.4 and 322.8~meV, respectively. These results agree with the latest experimental measurements \cite{Walter2010,Walter2020} within~1\%.
The significant disagreement between the present theoretical EAs and the (too large) LPES values~\cite{Williams1998,Carpenter2000}  for both In and Tl systems allows us to discard the latter values against the LPTS~\cite{Walter2010,Walter2020} results.
As far as the suggestion made by Felfli {\it et al.}~\cite{Felfli2012} is concerned, interpreting the previous thallium electron affinities as the binding energy of the first excited states of Tl$^-$,
the present MCDHF-RCI results firmly confirm that the
experimental value of 320.053(19)~meV~\cite{Walter2020} should be definitely  assigned to the real electron affinity.
The present calculations indeed definitely exclude the possibility of a more bound state  than $6s^22p^2 \; ^3P_0$.
Excitation ground configuration energies of In$^-$ and Tl$^-$ and their radiative lifetimes are also estimated.
We confirm that In$^-$ has three bound states $^3P_{0,1,2}$, while Tl$^-$ only has one bound state $^3P_{0}$.

The isotope shift on the EAs of along In and Tl isotopes are estimated using the currently available rms nuclear radii.
Although the MS contributes significantly to the indium EAs, it is already smaller than the FS contribution. For the EA(Tl), the FS largely dominate the MS that becomes almost negligible.
The isotope shift  between the two stable  Tl$^{203}$ and Tl$^{205}$ isotopes is estimated to be
${\rm IS(EA)}^{205}-{\rm IS(EA)}^{205}=+0.7013$~GHz, corresponding to a ``normal'' isotope shift due to the gain in electron density
at the nucleus accompanying the outer electron detachment from the anion.

The lifetimes estimated for the the excited In$^-$ fine structure $6s^26p^2 \; ^3P_{1,2}$ levels  are rather long but could be measured using a cryogenic ion storage ring. We hope that the present work will stimulate such experiments in that line.

\section*{Acknowledgement}
We acknowledge support from the Belgian FWO and FNRS Excellence of Science Programme (EOS-O022818F). SS is a FRIA grantee of the F.R.S.-FNRS. CYC acknowledges support from the National Natural Science Foundation of China (Grant No. 12074081 and 11974080).
KW expresses his gratitude to the support from the National Natural Science Foundation of China (Grant No.~11703004), and the Natural Science Foundation of Hebei Province, China (A2019201300).

\clearpage
\bibliographystyle{aip}
\bibliography{Tl}

\end{document}